# Electrical Detection of Light Helicity using a Quantum Dots based Hybrid Device at Zero Magnetic Field

F. Cadiz [a], D. Lagarde [a], B. Tao [b], J. Frougier [c], B. Xu [d], H. Jaffrès [c],
Z. Wang [d], X. Han [e], J. M. George [c], H. Carrere[a], A. Balocchi[a],
T. Amand [a], X. Marie [a], B. Urbaszek [a], Y. Lu [b] and P. Renucci [a]

[a] Université de Toulouse, INSA-CNRS-UPS, LPCNO, 135 Av. Rangueil, 31077 Toulouse, France
[b] Institut Jean Lamour, UMR 7198, CNRS- Nancy Université, BP 239, 54506 Vandoeuvre, France
[c] Unité Mixte de Physique CNRS/Thales and Université Paris-Sud 11, 1 Avenue Augustin Fresnel, 91767 Palaiseau, France
[d] Key Laboratory of Semiconductor Materials Science, Institute of Semiconductors, Chinese Academy of Sciences, P. O. Box 912, Beijing 100083, China
[e] Beijing National Laboratory of Condensed Matter Physics, Institute of Physics, Chinese Academy of Sciences, Beijing 100190, China

## ABSTRACT

Photon helicity-dependent photocurrent is measured at zero magnetic field on a device based on an ensemble of InGaAs/GaAs quantum dots that are embedded into a GaAs-based p-i-n diode. Our main goal is to take advantage of the long electron spin relaxation time expected in these nano-objects. In these experiments, no external magnetic field is required thanks to the use of an ultrathin magnetic CoFeB/MgO electrode, presenting perpendicular magnetic anisotropy (PMA). We observe a clear asymmetry of the photocurrent measured under respective right and left polarized light that follows the hysteresis of the magnetic layer. The amplitude of this asymmetry at zero magnetic field decreases with increasing temperatures and can be controlled with the bias. Polarization-resolved photoluminescence is detected in parallel while the device is operated as a photodetector. This demonstrates the multifunctional capabilities of the device and gives valuable insights into the spin relaxation of the electrons in the quantum dots.

## 1. INTRODUCTION

Intense investigations have been led on spin injection and spin properties in semiconductors in the last twenty years, in view of developing opto-spintronics devices [1]–[3] offering new functionalities based on the electron spin degree of freedom. In particular, spin Light Emitting Diodes (Spin-LED) containing ensembles of quantum dots [4]–[10], where spin polarized electrons are electrically injected, could be of particular importance for quantum information processing [11] (an electron spin confined in a semiconductor quantum dot can be used as a quantum bit) and quantum cryptography [12]. An efficient way to realize electrical spin injection into the dots consists in injecting spin-polarized electrons from a ferromagnetic injector via a tunnel barrier [13], [14] (e.*g*. an oxide barrier like MgO [15]–[22]) ; this allows to overcome the problem of conductivity mismatch between the metal and the semiconductor [23], [24]. Recently, results of spin injection processes, with electron spin aligned along the growth axis Oz of the device, was reported in quantum wells and quantum dots *without* any external magnetic field (that was previously required to rotate the magnetization of the electrode along Oz). This breakthrough was realized thanks to the use of an ultrathin CoFeB(/MgO) injector characterized by a strong perpendicular magnetic anisotropy (or PMA) [10], [18], [22].

Conversely, spin-optronics detectors or spin-photodiodes, for which the amplitude of the photocurrent depends on the helicity of the detected photon and on the direction of the magnetization of the ferromagnetic electrode, may be of interest for optical telecommunications based on photon polarization [25]. This new generation of devices could be also very useful for memory-based devices with an optical reading of information stored in the ferromagnetic layer. Several works have been published dealing with such helicity sensitive detectors based on metal-oxide-semiconductor (MOS) junctions [26]–[29] or Spin-LEDs used under reversed bias (through the paper, we define the qualification reversed bias with respect to the p-i-n junction, and not relative to the magnetic tunnel injector and ferromagnetic metallic electrode) [30]. The key issues to obtain an efficient helicity detector are: (i) an electron spin relaxation time in the semiconductor part long enough such that the spin is maintained until the carrier crosses the tunnel barrier before to be collected by the ferromagnetic electrode, (ii) the ability to work at zero magnetic field for a realistic device. Up to now, most of the studies have been performed using external magnetic fields of the order of 1 Tesla [27], [28], [30] to align the magnetization of the electrode along the growth direction of the structure, in order to obey the condition imposed by the optical selection rules in the semiconductor [31]. Among these studies, a few of them were realized at zero external magnetic field using PMA injectors [32] or in-plane injectors [33]–[35] under oblique incidence angle. (iii) and also important, the existence of a carrier recombination channel [30], [36] that is competitive with the extraction channel from the semiconductor to the ferromagnetic electrode, in order to get an helicity asymmetry of the photocurrent under circularly polarized light in continuous-wave (CW) experiments.

In this paper we propose to use p-doped quantum dots embedded in the active region of the detector in order to take benefit from the long electron spin relaxation time expected in such nano-objects [37], coupled with an ultrathin CoFeB magnetic electrode presenting PMA [38] to fulfill both (i), (ii) and (iii) requirements. Using such a hybrid system, we observe an asymmetry of photocurrent under right and left circularly polarized light (of about 0.4% at low temperature) that follows the hysteresis cycle of the magnetic layer. To the best of our knowledge, this is the first report of a spin detector based on semiconductor quantum dots. The photocurrent asymmetry amplitude decreases for increasing temperatures and can be controlled by the bias. Moreover, as emphasized in this paper, polarization-resolved photoluminescence is detected in parallel while the device is operated as a photodetector. This simultaneous measurement of both polarized photocurrent and photoluminescence give new insights into electron spin relaxation inside the dots when the bias is varied.

## 2. SAMPLE AND EXPERIMNTAL SET-UP

The schematics of the Spin-LED structure is displayed in Fig.1(a). The p-i-n LED device grown by MBE contains three planes of InGaAs quantum dots embedded in the intrinsic region. The full sequence of the semi-conductor structure is the following: p+-GaAs:Zn (001) substrate (p = $3 \times 10^{18}$ cm$^{-3}$ )/ 300 nm p-GaAs:Be (p = $5 \times 10^{18}$ cm$^{-3}$)/ 400 nm p-Al$_{0.3}$Ga$_{0.7}$As:Be (p = $5 \times 10^{17} - 5 \times 10^{18}$ cm$^{-3}$)/3*[30 nm undoped /7ML InGaAs QD (Be delta doping)]/ 30 nm undoped GaAs/ 50 nm n-GaAs:Si (n = $10^{16}$ cm$^{-3}$). The LED surface is passivated with arsenic in the semiconductor MBE chamber and then transferred through air into a second MBE-sputtering interconnected system. The As capping layer is firstly desorbed at 300°C in the MBE chamber and then the sample was transferred through ultra-high vacuum to a sputtering chamber to grow the MgO layer of thickness 2.5 nm. Finally, the 1.1 nm Co$_{0.4}$Fe$_{0.4}$B$_{0.2}$ spin injector and 5 nm Ta protection layer are deposited by sputtering in both cases. Concerning device fabrication, 300 μm diameter circular mesas were then processed using standard UV photolithography and etching techniques. Finally, the processed wafers were cut into small pieces to perform rapid temperature annealing (RTA) at 300°C for 3 minutes. Detailed growth conditions can be found elsewhere [18].

The sample was kept in a vibration-free closed-cycle He cryostat equipped with a superconducting coil enabling to apply magnetic fields up to ±2 T normal to the sample plane. The photocurrent measurements were performed under CW excitation with 850 nm and 910 nm wavelengths provided by laser diodes focused onto the sample with a typical spot diameter of 3 μm. The helicity of the laser diode is modulated by a photoelastic modulator (PEM) at 50 kHz. The difference between the two photocurrents under circularly right and left polarized light is labelled $\Delta I_{ph}$. $\Delta I_{ph}$ is measured thanks to a trans-impedance amplifier (gain $10^5$) followed by a lock-in amplifier. The bias applied to the structure is provided by a source-meter that also measures the average photocurrent $I_{ph}$. The positive bias corresponds to inject electrons from the metal part to the semiconductor part. The laser power is kept below 20 μW to stay in the linear regime. An original point of our set-up is that the photoluminescence (PL) can be detected at the same time as the photocurrent (see Fig. 1 (a)). The PL was detected in the Faraday geometry with an aspheric lens (f = 8mm, NA= 0.5). The collected light was dispersed by a spectrometer and recorded by a cooled CCD-camera. For electroluminescence (EL) experiments, the light collection system is the same as for PL, as well as the spectral and polarization-resolved set-up. The EL(PL) circular polarization degree $P_c$ was analyzed through a λ/4 wave plate and a linear analyzer. $P_c$ is defined as $P_c = (I^+ - I^-)/(I^+ + I^-)$ where $I^+$ and $I^-$ are the intensities of the right and left circularly polarized components of the EL/PL, respectively.

## 3. CHARACTERIZATION OF THE C$_O$F$_E$B/MGO/G$_A$A$_S$ INTERFACE BY ELECTROLUMINESCENCE AT ZERO MAGNETIC FIELD

The inset of Fig. 1(b) shows a typical polarization-resolved EL spectrum at 15 K and at zero magnetic field after the CoFeB electrode has been magnetized up to saturation along the z growth axis. Remarkably, the ensemble of quantum dot emission shows a significant circular polarization of $P_c$ = 18 % when a DC current of magnitude 160 μA is applied. Thanks to the p-doping of the dots (one hole per dot in average), the emitting quasi-particles are positively charged excitons (trions $X^+$), constituted by two valence holes and one conduction electron, which are circularly polarized, contrary to neutral excitons that are linearly polarized (due to the shape anisotropy of the quantum dots) [37], [39], [40]. The resulting optical selection rules are quite straightforward: $P_c$ is equal in first approximation to the electron's spin polarization $P_s$ ($P_s = (n^- - n^+)/(n^- + n^+)$, where $n^-$ and $n^+$ are respectively the number of electrons with spin down and up in the ensemble of quantum dots, provided that the electron spin relaxation time within the trion is longer than the electron lifetime (which is already validated for InGaAs p-doped quantum dots [37]). Long electron spin relaxation time (a few ns) in p-doped quantum dots [37] is due to the fact that the exchange interactions between the electron and the two holes with opposite spin cancel each other within the trion quasi-particle. Figure 1. (b) displays the EL circular polarization at 15 K as a function of the external magnetic field at a fixed bias of V = 2.2 V (I = 80 µA). The EL polarization follows the CoFeB magnetic hysteresis cycle. Remarkably we find $P_c$ ~20% at B=0. The ultrathin CoFeB layer presents a perpendicular magnetic anisotropy due to the interfacial anisotropy at the CoFeB/MgO interface [38], that is responsible for this PMA hysteresis. Note that the magnetic circular dichroism (MCD) is below 1% for this type of injector [18].

In Fig. 1 (b) small circular polarization dips are observed when magnetic field is closed to B = 0T. In the 0-100 mT range, this may be due to the electron spin relaxation induced by the hyperfine interaction between the nuclear spins and the electron spin localized in the dot [41]. In this picture, this spin relaxation mechanism is quenched when the external magnetic field is applied from 0 to about 100 mT [37], leading to an increase of $P_c$. For magnetic fields larger than 100mT, we observe a slight increase of $P_c$. This can result from the possible circularization of the neutral exciton eigenstates [40] (the broad luminescence may include a (small) fraction of neutral excitons X recombination, in addition to the $X^+$ trions component). Magnetic domains alignment in the magnetic layers close to the interface may also play a role in the slight $P_c$ increase measured in the 0.2 – 1 Tesla range. The precise physical origin of the shape of the polarization hysteresis loop will require further investigations. However the observation of an EL circular polarization of about 20% at B=0T is a clear manifestation of a very efficient electrical spin injection without external magnetic field, and proves the very high quality of the CoFeB/MgO/GaAs interface.

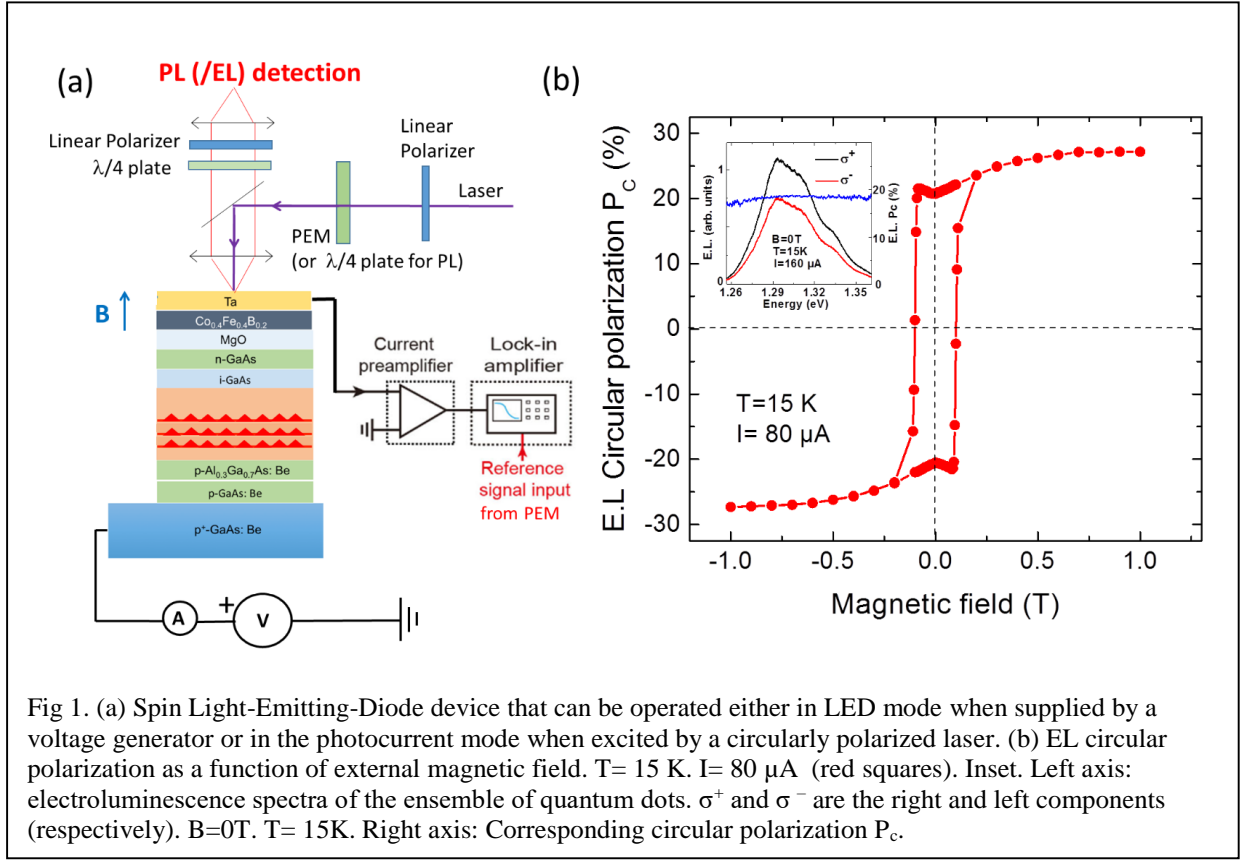

Fig 1. (a) Spin Light-Emitting-Diode device that can be operated either in LED mode when supplied by a voltage generator or in the photocurrent mode when excited by a circularly polarized laser. (b) EL circular polarization as a function of external magnetic field. T= 15 K. I= 80 µA (red squares). Inset. Left axis: electroluminescence spectra of the ensemble of quantum dots. $\sigma^+$ and $\sigma^-$ are the right and left components (respectively). B=0T. T= 15K. Right axis: Corresponding circular polarization $P_c$.

## 4. HELICITY DEPENDENT PHOTOCURRENT

We then turn to photocurrent experiments under circularly polarized light. The ensemble of quantum dots is excited by a laser diode at 910 nm (corresponding to high energy InGaAs dots emission/absorption) and 850 nm (close to the wetting layer states energy) with comparable results. Thanks to our specific set-up, it is possible to detect the photoluminescence signal at the same time as the photocurrent detection. Fig. 2 (a) displays both the photocurrent and PL signals as a function of the applied bias on the device. As expected, the two signals are complementary. When a positive bias is applied to the structure, such that the p-i-n junction is working under forward bias (but below the bias threshold required for the appearance of EL signal), the photocurrent intensity is decreasing while by correlation the PL signal is increasing. This is due to the fact that the bands are getting flatter when the bias increases [21]. On the contrary, for negative bias, the bending of the bands increases in the intrinsic region of the structure, and the extraction of carriers out of the active region by tunnel effect is more efficient due to the strong electric field, leading to an increase of the photocurrent signal. In the same time, the PL signal decreases due to this escape of carriers out of the dots, that constitutes an efficient non-radiative channel.

We define an asymmetry factor A= $\Delta I_{ph}/I_{ph}$ (where $I_{ph}$ is the average photocurrent) to quantify the helicity dependent photocurrent. In Fig. 2 (d), at V=+0.4 V, one can observe that the asymmetry factor A is about 0.4% at zero magnetic field and T= 30K after the CoFeB layer has been magnetized at saturation along the z growth axis. Moreover, the factor A follows the hysteresis cycle of the ultrathin CoFeB magnetic layer. The observed coercitive field of about 100mT is comparable to the one measured by EL in Fig. 1 (b). Remarkably, the helicity-dependent photocurrent can be modified by applying an external bias (Fig. 2. (b)). Note that this bias dependence of A allows us to rule out a possible artefact due to magnetic circular dichroism (a flat background independent of the applied bias that may be due to a small MCD has been properly subtracted). It appears that A increases with the applied bias, in the forward regime for the p-i-n junction (in the flat band regime slightly before EL emission starts). We can see in Fig. 1 (a) that the bias range (symbolized by vertical blue dashed lines) for which A is maximum corresponds to a relatively weak photocurrent amplitude and a large PL intensity. In this regime, radiative recombination in the quantum dots constitutes a strongly competing channel to photocurrent extraction.

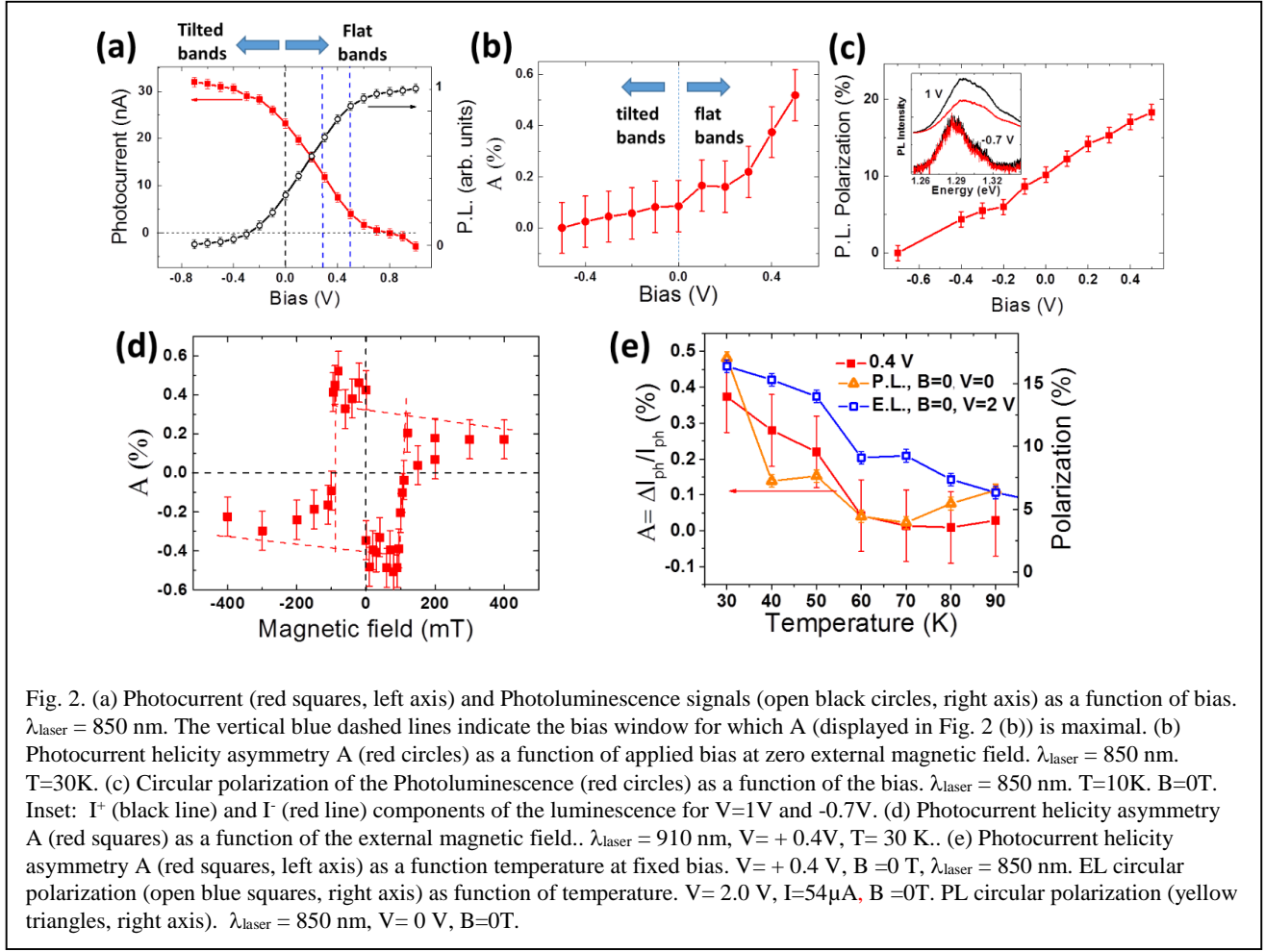

Fig. 2. (a) Photocurrent (red squares, left axis) and Photoluminescence signals (open black circles, right axis) as a function of bias. $\lambda_{laser}$ = 850 nm. The vertical blue dashed lines indicate the bias window for which A (displayed in Fig. 2 (b)) is maximal. (b) Photocurrent helicity asymmetry A (red circles) as a function of applied bias at zero external magnetic field. $\lambda_{laser}$ = 850 nm. T=30K. (c) Circular polarization of the Photoluminescence (red circles) as a function of the bias. $\lambda_{laser}$ = 850 nm. T=10K. B=0T. Inset: I$^+$ (black line) and I$^-$ (red line) components of the luminescence for V=1V and -0.7V. (d) Photocurrent helicity asymmetry A (red squares) as a function of the external magnetic field.. $\lambda_{laser}$ = 910 nm, V= + 0.4V, T= 30 K.. (e) Photocurrent helicity asymmetry A (red squares, left axis) as a function temperature at fixed bias. V= + 0.4 V, B =0 T, $\lambda_{laser}$ = 850 nm. EL circular polarization (open blue squares, right axis) as function of temperature. V= 2.0 V, I=54µA, B =0T. PL circular polarization (yellow triangles, right axis). $\lambda_{laser}$ = 850 nm, V= 0 V, B=0T.

Two main reasons could explain the large increase of A under forward bias. First, a channel competing with respect to the extraction of carriers from the dots (or localized states in the wetting layer) towards the magnetic electrode is required to obtain a sizeable photocurrent contrast [30], [36]. Indeed in the absence of any competing channel (PL, interface recombination …), CW excitation would lead to helicity-dependent spin-accumulation at the interface with the MgO barrier but, however, zero contrast of photocurrent. In our case, the radiative recombination channel in the dots could constitute a competing channel that becomes more efficient under forward bias, and would lead to a larger value of the photocurrent asymmetry.

The second possible reason is that the electron spin relaxation in the dots could depend on the applied bias. Some insight about this dependence can be obtained by polarization-resolved PL experiments under bias (Fig. 2 (c)). As expected, the spectra at -0.7V is red-shifted compared to the one at +1V, due to quantum confined Stark effect [42]. The spectrum observed at -0.7V is less broad at high energy, because of the easiest escape out of the dots of electrons by tunneling effect from high energy quantum dots under reverse bias. The PL circular polarization of the ensemble of dots (integrated on the whole spectrum) increases strongly from about 4% to about 18 % when the bias varies from -0.6V to +0.6V. We believe that this can be explained by the bias-induced change of trion/exciton population ratio within the ensemble of dots. Indeed a positive bias favors the presence of one resident hole in the dot, and thus the photogeneration of trions (which are circularly-polarized) instead of neutral excitons (whose eigenstates are linearly-polarized). The corresponding change in the trion/exciton population ratio yields an increase of the PL circular polarization. This polarization increase will also contribute to the larger value of the photocurrent asymmetry obtained at positive bias (Fig. 2(b)).

Finally, we have also studied the dependence of the helicity dependent photocurrent A= $\Delta I_{ph}$ /$I_{ph}$ as a function temperature (Fig. 2 (e)). A clear drop is observed when the temperature varies from 30 to 90 K. Above 90K, the signal disappears. The trend is the same for the EL circular polarization and the PL circular polarization (Fig. 2 (e)). This behavior is probably due to a phonon induced spin relaxation mechanism, clearly evidenced in InAs/GaAs quantum dots [43], [44]. Note that this temperature dependence also allows us to rule out artificial effect coming from MCD.

## 5. CONCLUSION

Thanks to the use of an ultrathin magnetic layer of CoFeB that presents perpendicular magnetic anisotropy, and p-doped InAs/GaAs quantum dots, we have evidenced at zero external magnetic field an asymmetry of photocurrent under right and left polarized light that follows the hysteresis cycle of the magnetic electrode. The amplitude of this asymmetry at zero magnetic field is about 0.4 % at low temperature, and can be controlled by the bias. It paves the way for versatile devices operating at zero magnetic field with two functionalities (helicity dependent photodetector and circularly polarized light source). Note that the use of an ultrathin CoFeB electrode of a few atomic layers is a strong advantage compared to thicker stacks [32], [45], [46] which provide the required perpendicular magnetic anisotropy, because it minimizes the light absorption in the magnetic layer. This is a crucial point for the realization of sensitive photodetectors.

## ACKNOWLEDGMENTS

We thank M. Hehn for help to develop the growth of ultrathin CoFeB layer by sputtering. F.C, P.R. and H.C acknowledge the Grant NEXT No. ANR-10-LABX-0037 in the framework of the "Programme des Investissements d'Avenir.". This work was also supported by the joint French National Research Agency (ANR)-National Natural Science Foundation of China (NSFC) SISTER project (Grants No. ANR-11-IS10-0001 and No. NNSFC 61161130527) and ENSEMBLE project (Grants No. ANR-14-0028-01 and No. NNSFC 61411136001), and by the grant NEXT No ANR-10-LABX-0037 in the framework of the Programme des Investissements d'Avenir. Experiments were performed using equipment from the platform TUBEDavm funded by FEDER (EU), ANR, the Region Lorraine and Grand Nancy. X.M acknowledges the Institut Universitaire de France.